\documentclass[aps,prb,,showpacs,twocolumn,groupedaddress]{revtex4}

\usepackage{amsmath,bm,amsfonts,amssymb}
\usepackage{graphics,graphicx}

\begin{document}

\title{Microwave conductivity of ${\bm{d}}$-wave superconductors with extended impurities}

\author{Tamara S. Nunner}
\author{P.J. Hirschfeld}

\affiliation{Department of Physics, University of Florida,
Gainesville, FL  32611-8440, USA }

\date{\today}

\begin{abstract}
We investigate the influence of extended scatterers on the finite
temperature and finite frequency microwave conductivity of
$d$-wave superconductors. For this purpose we generalize a
previous treatment by Durst and Lee [Ref.~\onlinecite{DurstLee}],
which is based on a nodal approximation of the quasiparticle
excitations and scattering processes, and apply it to the analysis
of experimental spectra of YBCO-123 and BSCCO-2212. For YBCO, we
find that accounting for a slight spatial extension of the strong
scattering in-plane defects improves the fit of the low
temperature microwave conductivity to experiment.
With respect to BSCCO we conclude that it is necessary to  include
a large concentration of weak-to-intermediate strength extended
scatterers, which we attribute to the out-of plane disorder
introduced by doping.
These findings for BSCCO are consistent  with similar analyses of
the normal state ARPES spectra and of STM spectra in the
superconducting state, where an enhanced forward scattering has
been inferred as well.
\end{abstract}
\maketitle
\section{Introduction}
 Early  in the debate over  the symmetry of the
superconducting order in the  cuprates, a rather convincing
picture of the microwave properties of the YBCO-123 system was put
forward by Bonn {\it et al.}, \cite{bonnetal93} and later placed on a
microscopic foundation\cite{HPS,Rieck,SchachingerCarbotte}.
Crucial to this interpretation is the
observed collapse of the $d$-wave nodal quasiparticle scattering
rate as the system becomes
superconducting\cite{bonnetal91,nussetal,romeroetal},
leading to a
dramatic rise in the conductivity with decreasing temperature.
This rise is cut off when the inelastic mean free path becomes
comparable to the elastic one, and the conductivity subsequently
decreases because of the vanishing  nodal carrier density as
temperature tends to zero. One consequence of this picture is that 
the resulting conductivity peak should be suppressed 
 and occur at higher temperatures in dirtier samples.
In addition, the conductivity should approach
the universal value $\sigma_{0}\equiv e^2v_F/h v_2$ for zero
temperature and zero frequency as predicted by P.A. Lee\cite{PALee93} for
the case of isotropic scatterers, where 
$v_F$ is the Fermi velocity and $v_2$ the gap velocity
at the node.

If one extracts $v_F/v_2$ from thermal conductivity\cite{TailleferDiraccone}
or angle resolved photoemission (ARPES) measurements\cite{ARPESDiraccone},
one finds that the universal value for both BSCCO and YBCO crystals should be
about $\sigma_0=5\times 10^5 \Omega^{-1}{\rm m}^{-1}$. 
In YBCO, the residual value of the conductivity for $\omega$,$T\to$0
is difficult to determine, but appears to be approaching 3-4$\times \sigma_{0}$ in the best
crystals\cite{hosseinietal}. The peak in the conductivity occurs around
25K with an amplitude of approximately 100$\times \sigma_0$ for the lowest
frequency measured.
In BSCCO, the peak is located around 20K, but is only about 20\%
higher than the apparent residual value~\cite{leeetal}
of  8-10$\times \sigma_{0}$. Virtually no frequency dependence 
is seen in the measured microwave frequency range~\cite{leeetal}, 
suggesting a very dirty material, in apparent
contradiction--within the ``standard" scenario--with the
low-temperature peak position. The longstanding puzzle of the low
temperature microwave peak together with indications of dirty
limit behavior have been analyzed  as evidence for absorption into
a collective mode off resonance at low
frequencies\cite{orenstein1}, as well as a consequence of
nanoscale inhomogeneity\cite{orenstein2}.

In this work we argue that the temperature dependence of the
conductivity can be more naturally understood in terms of the
effect of extended scatterers present in the BSCCO crystal.
Current generation crystals are made typically with excess Bi,
deficiencies of Sr and Ca, and variable O content; cation
substitution is thought to occur frequently. Some aspects of this
defect distribution have been discussed recently by Eisaki 
{\it et al.}\cite{Eisaki}. The net result of these defects is not only to
dope the nominally stoichiometric BSCCO crystal (pure BSCCO would
be an insulator), but to provide a relatively slowly varying
potential landscape for quasiparticles moving in the CuO$_2$
planes.
The effect of these extended scatterers with respect to the normal state 
has recently been discussed by Abrahams and Varma\cite{AV00} 
and it has been pointed out by Zhu {\it et al.}\cite{ZHA} 
that the broad momentum space peaks observed in Fourier transform
STM studies of BSCCO\cite{Howaldetal,Hoffmanetal,McElroyetal} can
only be explained by potential scatterers with finite range.  In a
further application of this notion to ARPES, Zhu {\it et
al.}\cite{ZHS} showed that a large concentration of impurities with potentials
peaked in the forward direction could be present without
substantially
broadening quasiparticle states except near the node. Since the
microwave conductivity is dominated by nodal quasiparticles, it is
clearly important to ask what the effects of extended or forward
scatterers are in this case.

Since the work of Durst and Lee,\cite{DurstLee} we know  that the
residual conductivity in the presence of extended scatterers can
be much larger than the ``universal" value $\sigma_{0}$.  This
might  account for the large value of the microwave conductivity
observed in the BSCCO-2212 system at low temperatures.
To make this case, however, one needs to
examine the influence of a finite scattering range at finite
temperatures and frequencies. We have therefore generalized the
analysis of Durst and Lee in this way
and applied this treatment to the analysis of
experiments on YBCO and BSCCO. With respect to YBCO we find
that consideration of slightly extended strong scatterers provides
a better fit to the low temperature microwave conductivity than
pointlike strong scatterers. For BSCCO we conclude that it
is necessary to include a large concentration of weak-to-intermediate strength
extended impurities in addition to the strong in-plane
defects which are responsible for the unitary scattering resonances
observed by STM~\cite{Hudson}.

The outline of the paper is at follows.
In section~\ref{sec:Model}, we describe the model and derive
expressions for the self-energy and vertex function for
extended scatterers. Our approach, which is based on an extension of the work
by Durst and Lee~\cite{DurstLee}, aims at treating scattering
potentials with an extension of a few lattice spacings at maximum and
is therefore in the opposite limit from
semiclassical calculations where the impurity
potentials extend over a few coherence lengths~\cite{Adagideli,Sheehy}.
In section~\ref{sec:YBCO} we apply our treatment
to the microwave conductivity of YBCO. We show that
the consideration of slightly extended instead of pointlike
strong potential scatterers improves the agreement with the
experimental spectra. In section~\ref{sec:BSCCO} we address the
microwave conductivity of BSSCO and demonstrate that it is necessary
to include a large concentration of weak extended scatterers
in order to explain the experimental spectra. A good fit is obtained
based on a realistic disorder model for
BSCCO which contains weak extended scatterers in addition to strong
pointlike impurities. Finally, in section IV, we present our
conclusions.

\section{Treatment of extended scatterers}
\label{sec:Model}

For low temperatures and low frequencies the quasiparticle
dispersion of a $d$-wave superconductor can be linearized around
the nodes. The resulting quasiparticle spectrum has the form of a
Dirac cone, whose anisotropy is determined by the ratio $v_f/v_2$
of the Fermi velocity $v_f=\partial \epsilon_k/ \partial
k=2\sqrt{2}t$ and the gap velocity $v_2=\partial \Delta_k/\partial
k=\Delta_0/\sqrt{2}$, where $t$ is the nearest neighbor hopping
parameter and $\Delta_0$ is the maximum gap value and we have set
$a=\hbar=1$. At low temperatures and frequencies, quasiparticle
excitations are restricted to small regions around the nodes.
Therefore momentum transfer between quasiparticles is limited to
four wavevectors which connect the four nodes and include
intranode and internode scattering processes (see
Fig.~\ref{fig:NodalApprox}). Consequently a momentum dependent
impurity potential $V(k)$ can be represented by three parameters
$V_1$, $V_2$ and $V_3$ which correspond to the respective momentum
transfers. Based on these approximations and treating impurity
scattering in T-matrix approximation, an expression for the
microwave conductivity has been derived by Durst and
Lee~\cite{DurstLee}. They found that vertex corrections, which
arise from the momentum dependence of the impurity potential,
induce a dependence of the zero-temperature and zero-frequency
limit of the conductivity on the impurity potential and the
impurity concentration. Contrary to the case of pointlike
scatterers, no universal value of the electrical conductivity
exists therefore in case of extended scatterers. Durst and Lee,
however, did not further explore the frequency and temperature
dependence of the microwave conductivity. Here, we slightly modify
their approach and evaluate the conductivity for finite
frequencies and temperatures. Furthermore we consider the effect
of combining different types of scatterers and calculate the
microwave conductivity for BSCCO based on a realistic disorder
model.

\subsection{Self-energy}
\label{sec:SelfEnergy}

\begin{figure}[t]
\begin{center}
\leavevmode
\includegraphics[clip=true,width=.45\columnwidth]{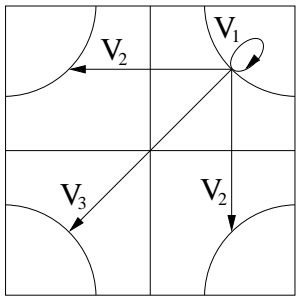}
\caption{For low temperatures and low frequencies the momentum
transfer between quasiparticles is basically limited to the four
wavevectors connecting the nodes of a $d$-wave
superconductor. Therefore the momentum dependent impurity potential
$V(k)$ can be represented by the values at the respective wavevectors,
i.e., by three parameters $V_1$, $V_2$ and $V_3$.}
\label{fig:NodalApprox}
\end{center}
\end{figure}

Before proceeding to two-particle quantities like the microwave
conductivity, it is instructive to focus first on single-particle
properties like the single-particle self-energy.
Using the Nambu notation, the single-particle self energy in a
superconductor can be decomposed as:
\begin{equation}
\tilde \Sigma(k,\omega)= \sum_\alpha \Sigma_\alpha(k,\omega) \tilde \tau_\alpha \,,
\end{equation}
where $\tilde \tau_\alpha$ are the Pauli matrices and $\tilde \tau_0$ is the unit matrix.
Treating impurity-scattering in T-matrix approximation gives rise to
the following self-energy
\begin{equation}
\tilde \Sigma(k,\omega)= n_i \tilde T_{kk}(\omega),
\end{equation}
where $n_i$ is the impurity concentration and $T_{kk}(\omega)$ is the
diagonal element of the T-matrix
\begin{equation}
\label{eq:TMatrixk}
\tilde T_{kk'}(\omega)= V_{kk'} \tilde \tau_3
+ \sum_{k''} V_{kk''} \tilde \tau_3 \tilde G(k'',\omega) \tilde T_{k''k'}(\omega)
\,.
\end{equation}
The self-energy $\tilde \Sigma(k,\omega)$ has to be solved
self-consistently in combination with the single-particle Green's function
\begin{equation}
\tilde G(k,\omega)^{-1}=\tilde G_0(k,\omega)^{-1}
-\tilde \Sigma(k,\omega)
\end{equation}
where the unperturbed Green's function is given as
\begin{equation}
\tilde G_0(k,\omega)=
\frac{\omega \tilde \tau_0 + \Delta_k \tilde \tau_1 +\epsilon_k \tilde
\tau_3}
{\omega^2-\epsilon_k^2-\Delta_k^2}\,.
\end{equation}
Following the approach of Durst and Lee~\cite{DurstLee}, we reduce the impurity
scattering potential to the four wave-vectors connecting
the nodes, i.e., $V_{kk'}$ is replaced by a $4\times4$-matrix in
nodal space
\begin{equation}
V_{kk'} \rightarrow \underline{V}=
\left( \begin{array}{cccc}
V_1 & V_2 & V_3 & V_2 \\
V_2 & V_1 & V_2 & V_3 \\
V_3 & V_2 & V_1 & V_2 \\
V_2 & V_3 & V_2 & V_1
\end{array} \right) \,,
\end{equation}
where $V_1$, $V_2$ and $V_3$ are the values of the impurity potential
at the wavevectors connecting the nodes, see Fig.~\ref{fig:NodalApprox}.

\begin{figure}[t]
\leavevmode
\begin{center}
\includegraphics[clip=true,width=.8\columnwidth]{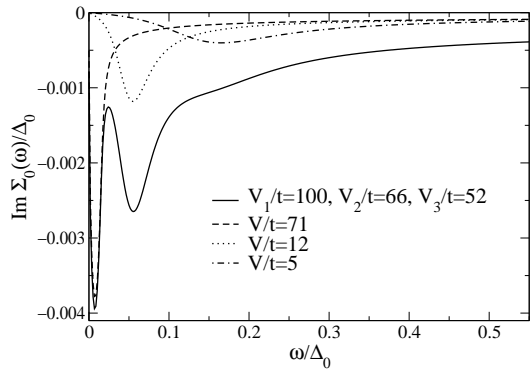}
\caption{Imaginary part of the self-consistently calculated self-energy
$\Sigma_0(\omega)$ for an
anisotropic impurity potential characterized by the three parameters
$V_1/t=100$, $V_2/t=66$ and $V_3/t=52$ (solid line). The positions of
the resonances coincide with the resonances of
the self-energies for isotropic impurity potentials $V/t=71$,
$V/t=12$ and $V/t=5$. Here $\Delta_0/t=.29$ and $n_i=.00002$ have been used.}
\label{fig:SelfEn3Peaks}
\end{center}
\end{figure}

Using this simplification, the impurity potential can be
pulled out of the integral and the T-matrix becomes a
$4\times4$-matrix in nodal space
\begin{equation}
\tilde T_{jj'}(\omega)= V_{jj'} \tilde \tau_3
+ \tilde I_G(\omega) \tilde \tau_3 \sum_{j''} V_{jj''} \tilde T_{j''j'}(\omega),
\end{equation}
where $I_G(\omega)$ is the integral of the single-particle Green's
function over one quarter of the Brillouin zone
\begin{eqnarray}
\label{eq:SingleParticleGreen}
\tilde I_G(\omega) &=&  \int_{0}^{\pi} \int_{0}^{\pi}
\frac{d^2k}{(2\pi)^2} \tilde G (k,\omega)
\approx  I_G(\omega) \tilde \tau_0 \,.
\end{eqnarray}
This allows for an analytical solution~\cite{DurstLee} of the T-matrix
\begin{equation}
\tilde T_{jj'}= T_{jj'}^3 \tilde \tau_3 + T_{jj'}^0 \tilde \tau_0
\end{equation}
with
\begin{equation}
T_{jj'}^3=\left(\frac{\underline{V}}{1-I_G(\omega)^2\underline{V}^2}
        \right)_{jj'} ,\,\,
T_{jj'}^0=\left(\frac{-I_G(\omega)\underline{V}^2}{1-I_G(\omega)^2\underline{V}^2}
        \right)_{jj'} \,,
\end{equation}
where the denominators have to be calculated as inverse matrices
in nodal space.
This gives for the $\Sigma_0$-component of the self-energy:
\begin{eqnarray}
\label{eq:SelfEnT}
&&\!\!\!\!\!\!\!\Sigma_0(\omega)=-\frac{n_i}{4I_G(\omega)}
\Bigl(4  - \frac{2}{1\!-\!I^2_G(\omega)(V_1-V_3)^2}  \\
&& \!\!\!\!\!\!\! - \frac{1}{1\!-\!I^2_G(\omega)(V_1-2V_2+V_3)^2}
        - \frac{1}{1\!-\!I_G^2(\omega)(V_1+2V_2+V_3)^2} \Bigr) \nonumber
\end{eqnarray}
For an isotropic impurity potential, i.e., $V_1$=$V_2$=$V_3$=$V$ this
expression simplifies to:
\begin{equation}
\label{eq:SelfEnIso}
\Sigma_{0}^{\rm iso}(V,\omega)=-\frac{n_i}{4I_G(\omega)}
\Bigl( 1-\frac{1}{1-16 I_G^2(\omega) V^2} \Bigr) \,,
\end{equation}
which implies that the self-energy for a momentum-dependent impurity potential
in Eq.~(\ref{eq:SelfEnT}) can be decomposed into a sum of
self-energies corresponding to three different isotropic impurity
potentials
\begin{eqnarray}
\label{eq:SelfEnSum}
&&\Sigma_0(\omega)=
\Sigma_{0}^{\rm iso}(\frac{1}{4}(V_1+2V_2+V_3),\omega) \\
&&+\Sigma_{0}^{\rm iso}(\frac{1}{4}(V_1-2V_2+V_3),\omega)
+ 2 \Sigma_{0}^{\rm iso}(\frac{1}{4}(V_1-V_3),\omega) \,. \nonumber
\end{eqnarray}
Consequently, the self-energy for an anisotropic impurity potential in
the nodal approximation contains up to three resonances
corresponding to the different impurity strengths
$V=(V_1+2V_2+V_3)/4$, $V=(V_1-2V_2+V_3)/4$ and $V=(V_1-V_3)/4$, see
Fig.~\ref{fig:SelfEn3Peaks}.

If the scattering strength of the impurities is weak, they can
be treated within Born approximation and the self-energy of an
extended weak impurity becomes:
\begin{equation}
\label{eq:SelfEnBorn}
\Sigma_0(\omega)=-n_i (V_1^2+2V_2^2+V_3^2) I_G(\omega) \,,
\end{equation}
i.e., the self-energy for an anisotropic impurity potential is
identical to the self-energy for an isotropic impurity potential with
$V=\sqrt{V_1^2+2V_2^2+V_3^2}$.

\subsection{Microwave conductivity}

In linear response the electrical conductivity is given as:
\begin{equation}
\sigma (\Omega,T)=-\frac{{\rm Im} \Pi_{\rm
ret}(\Omega,T)}{\Omega}\,,
\end{equation}
where $\Pi_{\rm ret}(\Omega,T)$ is the retarded current-current
correlation function, which can be obtained by analytical continuation
from
\begin{equation}
\Pi(i\Omega)=\frac{e^2v_f^2}{\beta}\sum_{i\omega,k}
{\rm Tr}[\tilde G(k,i\omega) \tilde G(k,i\omega+i\Omega)
\tilde \Gamma(k,i\omega+i\Omega)] \,,
\end{equation}
where $\tilde \Gamma(k,i\omega+i\Omega)$ is the vertex function, which
for a $d$-wave superconductor arises entirely
from the momentum dependence of the impurity potential\cite{HWE88} and will be
calculated here as the sum of ladder diagrams.

\begin{figure}[h]
\begin{center}
\leavevmode
\includegraphics[clip=true,width=.9\columnwidth]{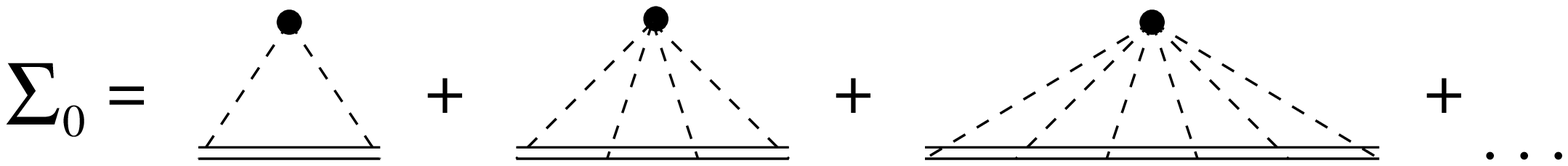}
\\[.5cm]
\includegraphics[clip=true,width=.9\columnwidth]{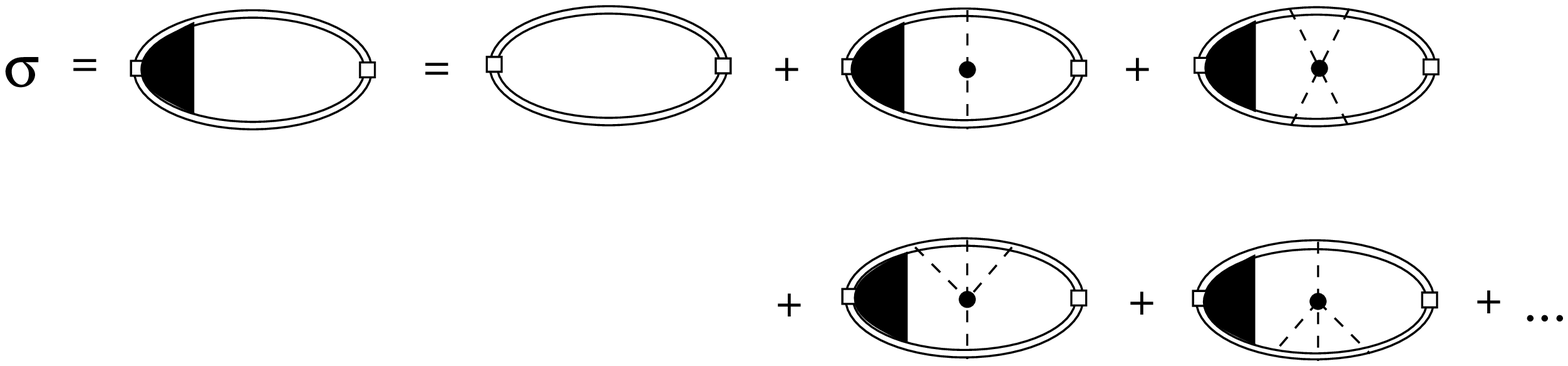}
\caption{T-matrix approximation for the self-energy $\Sigma_0$ and the
microwave conductivity $\sigma$. Because we neglect the
$\tilde \tau_3$-component of the self-energy, which corresponds to
neglecting all diagrams with odd number of impurity lines, all
diagrams with odd number of impurity lines have to be excluded from
the vertex corrections as well.}
\label{fig:TmatrixEven}
\end{center}
\end{figure}

For simplicity we neglect the $\Sigma_3$-component of the
self-energy and keep only the $\Sigma_0$-component. This
approximation becomes exact in the unitary limit for isotropic
scattering $V_1=V_2=V_3$ and for purely forward scattering
$V_2=V_3=0$ because the $\Sigma_3$-component vanishes in theses
limits and also in the Born approximation for all scattering
potentials. In terms of diagrams this means that all summations
are reduced to diagrams with even number of impurity lines. In
order to obtain a conserving approximation~\cite{BaymKadanoff} for
the conductivity,  the vertex corrections also have to be
restricted to diagrams containing even numbers of impurity lines,
see Fig.~\ref{fig:TmatrixEven}. This is a modification to the
treatment by Durst and Lee~\cite{DurstLee}, who have summed all
ladder diagrams both with even and odd number of impurity lines.
This modification is necessary when $\Sigma_3$ is neglected
because otherwise the vertex corrections violate analyticity at
finite $\omega$ and $T$. Furthermore, this modification leads to a
simplification to the expression of the conductivity obtained by
Durst and Lee~\cite{DurstLee}, see below. Note, however,  that our
treatment still agrees with Durst and Lee~\cite{DurstLee} in the
Born approximation, which is not affected by this modification and
also in the T-matrix approximation for the limit $\omega \to 0$
and $T \to 0$, i.e., the limit Durst and Lee have focused on.
Differences arise only for finite temperatures and frequencies in
the T-matrix approximation.

Summing up all ladder diagrams with even number of impurity
lines one arrives at the following expression for the current-current
correlation function:
\begin{equation}
\Pi(i\Omega)=\frac{e^2 v_f}{\pi v_2} \sum_{i\omega}
J(i\omega,i\Omega)
\end{equation}
with
\begin{equation}
\label{eq:J}
J(i\omega,i\Omega)=\frac{I_B(i\omega,i\Omega)}
         {1-\gamma(i\omega,i\Omega)I_B(i\omega,i\Omega)}
\end{equation}
where $I_B(i\omega,i\Omega)$ is the momentum integrated particle-hole
bubble
\begin{equation}
I_B(i\omega,i\Omega) =  \int_{0}^{\pi} \int_{0}^{\pi}
\frac{d^2k}{(2\pi)^2} \tilde G(k,i\omega) \tilde G(k,i\omega+i\Omega) \,.
\end{equation}
Note, that here we calculate all integrals $I_B(i\omega,i\Omega)$ and
$I_G(\omega)$ numerically instead of replacing the elliptical
integration area by a circle as in Ref.~\onlinecite{DurstLee}, because
this approximation becomes inaccurate for small values of the gap,
i.e., close to $T_c$, where we parameterize the  temperature dependence
of the gap in the following way:~\cite{DSH01}
\begin{equation}
\Delta_0(T)=\Delta_0 \tanh (\alpha \sqrt{T_c/T-1})
\end{equation}
with $\alpha=3.0$.

The vertex function is given by
\begin{eqnarray}
&\gamma(i\omega,i\Omega)&=\frac{n_i}{4\pi v_f v_2} \times\\
&&(T_{11}^0(i\omega)T_{11}^0(i\omega+i\Omega)
+T_{11}^3(i\omega)T_{11}^3(i\omega+i\Omega) \nonumber \\
&&\!\!-T_{13}^0(i\omega)T_{13}^0(i\omega+i\Omega)
-T_{13}^3(i\omega)T_{13}^3(i\omega+i\Omega))\,. \nonumber
\end{eqnarray}
After analytical continuation the microwave conductivity can be
expressed as~\cite{DurstLee}
\begin{eqnarray}
\label{eq:Conductivity}
&&\sigma(\Omega)=\frac{e^2 v_f}{2 \pi^2 v_2}
\int d \omega \frac{n_F(\omega)-n_F(\omega+\Omega)}{\Omega}
\\
&&\bigl( {\rm Re} J(\omega-i\delta,\omega+\Omega+i\delta)
      -{\rm Re} J(\omega+i\delta,\omega+\Omega+i\delta)
\bigr) \,. \nonumber
\end{eqnarray}

In Born approximation one arrives at a very similar expression for the
conductivity, but the vertex function $\gamma (i\omega, i\Omega)$ in
Eq.~(\ref{eq:J}) is replaced by the simpler from
\begin{equation}
\label{eq:VertexBorn}
\gamma^{\rm Born}=\frac{n_i}{4\pi v_f v_2} (V_1^2-V_3^2) \,.
\end{equation}

So far we have focused on the effect of impurity scattering,
which is appropriate for low temperatures and low
frequencies. At higher temperatures, however, it
is essential to take into account inelastic
scattering processes as well, like e.g. quasiparticle quasiparticle scattering or
scattering off spin fluctuations. These inelastic
scattering processes are suppressed below $T_c$ due to the opening of
the superconducting gap in the quasiparticle excitation spectrum and
therefore the contribution of inelastic scattering increases rapidly
when $T_c$ is approached from the low temperature side. A full treatment of
inelastic scattering is beyond the scope of this paper.
It has been pointed out by Walker and Smith~\cite{WS00}
that the contribution of quasiparticle quasiparticle scattering to the transport
lifetime is exponentially suppressed for low temperatures because only
Umklapp scattering processes can decay the current and a
finite excitation energy $\Delta_U$ is necessary to permit an
Umklapp scattering process for a realistic Fermi surface.
Thus, we include the effect of inelastic scattering by simply 
adding the inverse transport lifetime $\tau_{\rm inel}^{-1}(T)$, which
has been obtained by Duffy {\it et al.}~\cite{DSH01} by extracting
the Umklapp scattering processes from scattering of
quasiparticles off spin fluctuations,
to the imaginary part of the self-energy $\Sigma_0(\omega)$ in
Eq.~(\ref{eq:SelfEnT}) or Eq.~(\ref{eq:SelfEnBorn})
\begin{equation}
\label{eq:SigmaInel}
\Sigma(\omega)=\Sigma_0(\omega)-i (2 \tau_{\rm inel}(T))^{-1}
\end{equation}
Note, that our simplified treatment of $\tau_{\rm
inel}^{-1}$ completely neglects the frequency dependence of
inelastic scattering and therefore limits our approach to small
frequencies in the microwave regime, and prevents us from
calculating the conductivity in the THz-range.

\section{Comparison with experimental spectra of YBCO}
\label{sec:YBCO}

\begin{figure}[t]
\begin{center}
\leavevmode
\includegraphics[clip=true,width=.9\columnwidth]{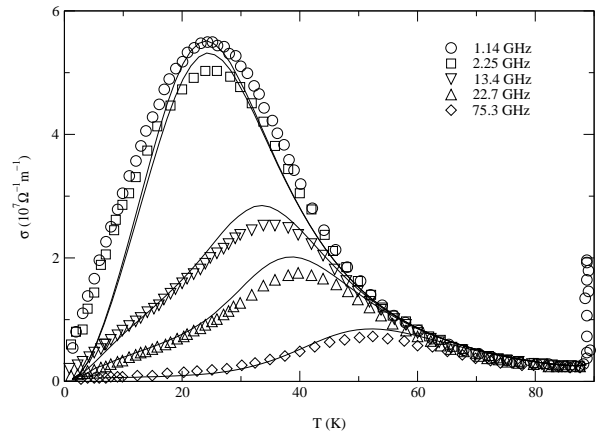}
\caption{Fit to the experimental spectra of YBCO (reproduced from
Ref.~\onlinecite{hosseinietal}) using pointlike strong scatterers with $V=100t$
and a concentration of $n_i=.000035$. 
The magnitude of the conductivity has been scaled by a factor of .42,
which corresponds to assuming a penetration depth of 1200\AA.}
\label{fig:FitYBCOiso}
\end{center}
\end{figure}

The microwave conductivity has been investigated in detail for
pointlike scatterers within the self-consistent T-matrix approximation~\cite{HPS}
and good agreement with the experimental data of YBCO has been found.
The temperature dependence of the
microwave conductivity for pointlike
scatterers, see also Fig.~\ref{fig:FitYBCOiso},
can be summarized in the following way. For zero
temperature and zero frequency the conductivity approaches a universal
value~\cite{PALee93}  $\sigma_0=e^2 v_f/h v_2$
due to the fact that at zero temperature
impurities give rise to a finite quasiparticle
density of states while at the same time they reduce the quasiparticle
lifetime. At low temperatures the conductivity rises with increasing $T$
due to an increase in the number of excited
quasiparticles.  The exact temperature dependence is
determined by the density of states in a $d$-wave superconductor and
the frequency dependence of the impurity
scattering rate. Starting from the opposite side, i.e., decreasing the
temperature below $T_c$ the conductivity also increases rapidly because inelastic
scattering is suppressed below $T_c$  due to the opening
of the superconducting gap in the quasiparticle excitation
spectrum. This leads to the formation of a peak
at intermediate temperatures, whose position is
determined by the microwave frequency
and the impurity scattering strength. This peak moves
to higher temperatures and its amplitude decreases with
increasing microwave frequency and impurity concentration.

Our best fit to the experimental spectra of
YBCO~\cite{hosseinietal} using pointlike strong scatterers is
displayed in Fig.~\ref{fig:FitYBCOiso}. Commonly used parameters
for YBCO are: $\Delta_0=400K$ for the gap maximum,  $v_f/v_2=14$
for the anisotropy of the Dirac cone and $T_c=88.7K$
\cite{TailleferDiraccone,hosseinietal}.
 In order to compare our theoretical curves to the
experimental data the value of the universal conductivity
$\sigma_0=e^2/(\hbar \pi^2) v_f/v_2$ has to be translated into a
three dimensional conductivity which can be done via
$\sigma_0^{3D}=\sigma_0^{2D} n_c$, where $n_c$ is the number of
CuO$_2$-planes per unit length in the c-direction, which is
$n_c$=1/(5.9\AA)  for YBCO.
Because $\sigma \sim \lambda^{-3}$ the conductivity $\sigma$ is
very sensitive to the value of the penetration depth $\lambda$,
which has recently been measured~\cite{Pereg-Barnea} as considerably
smaller than previously published in the literature~\cite{Tallon},
i.e., $\lambda=1030 \pm 80$\AA$\,$  instead of $\lambda\simeq 1550$\AA. 
This would increase the published values~\cite{hosseinietal} of the
microwave conductivity by a factor of four.
Indeed, it turns out that we obtain the best fit to the microwave
conductivity of YBCO when we assume the absolute values of the
conductivity to be roughly twice the previously published
data~\cite{hosseinietal}, which would correspond to a penetration
depth of approx. 1200\AA. Therefore we allow ourselves the freedom to
scale our calculated 
curves by roughly a factor of 1/2 when comparing to the
experimental published values (exact scaling factor is stated in
the figure captions).

\begin{figure}[t]
\begin{center}
\leavevmode
\includegraphics[clip=true,width=.9\columnwidth]{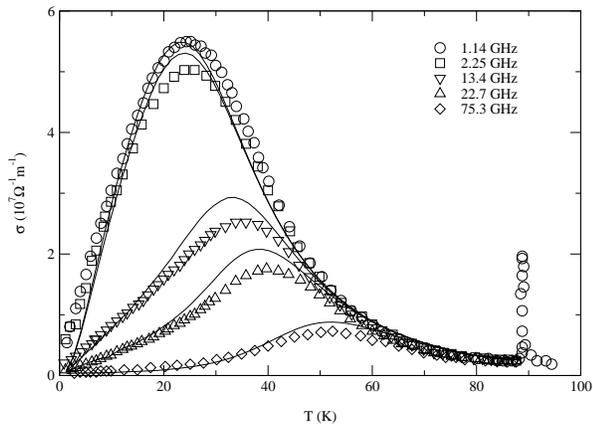}
\caption{Fit to the experimental spectra of YBCO (reproduced from
Ref.~\onlinecite{hosseinietal}) using slightly extended strong scatterers with
$V_1=100t$, $V_2=92t$, $V_3=82t$
and a concentration of $n_i=.000015$. 
The magnitude of the conductivity has been scaled by
a factor of .425 which corresponds to assuming a penetration depth of 1200\AA.}
\label{fig:FitYBCOext}
\end{center}
\end{figure}

As can be seen from Fig.~\ref{fig:FitYBCOiso} the assumption of
pointlike scatterers can reproduce the temperature and frequency
dependence of the microwave conductivity of YBCO quite well
(see Refs.~\onlinecite{HPS,Rieck,SchachingerCarbotte}). The
largest discrepancy arises for low temperatures and low frequencies,
where experimentally a nearly linear increase of the conductivity with
temperature is found whereas the theory based on pointlike scatterers
predicts a quadratic temperature dependence. It has been suggested by Hettler
and Hirschfeld~\cite{Hettler} that the theoretical lineshape becomes
more linear at low temperatures when the suppression of the order
parameter surrounding a strong pointlike scatterer is taken into
account. This low temperature behavior has been attributed to the
formation of a second resonance in the self-energy $\Sigma_0$ at low
frequencies. It is intriguing to note that we find a similar resonance in the
self-energy for slightly extended strong potential scatterers, see
Fig.~\ref{fig:SelfEn3Peaks}, and therefore it is interesting to investigate
whether the presence of slightly extended potential scatterers can also explain
the linear $T$-dependence of the microwave conductivity for low
temperatures. Our best fit to the experimental spectra of YBCO using
slightly extended potential scatterers is displayed in
Fig.~\ref{fig:FitYBCOext}.
Obviously the consideration of extended scatterers considerably
improves the agreement with the experimental data at low
temperatures.

\begin{figure}[b]
\begin{center}
\leavevmode
\includegraphics[clip=true,width=.98\columnwidth]{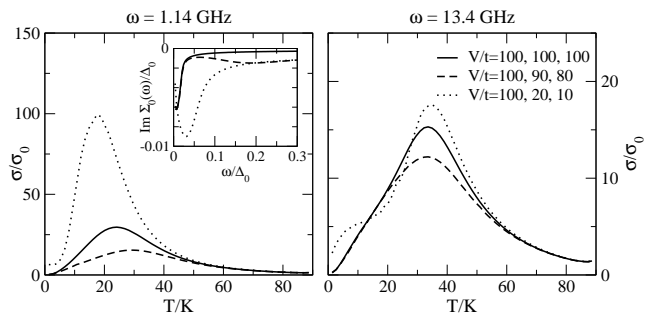}
\caption{Microwave conductivity for strong scatterers with varying
degree of forward scattering. Impurity concentration:
$n_i=.000035$, maximum gap: $\Delta_0=.29t$. Left panel:
$\omega=1.14$GHz, right panel: $\omega=13.4$GHz. Inset:
Self-energy for the respective scattering potentials.}
\label{fig:OptCondStrongImpV}
\end{center}
\end{figure}

Surprisingly, the concentration of extended scatterers used for
the fit in Fig.~\ref{fig:FitYBCOext} is even lower than the
concentration of pointlike scatterers we have used for the fit in
Fig.~\ref{fig:FitYBCOiso}. Generally one would assume that due to
the forward scattering character of extended impurities a larger
concentration is necessary to obtain a similar scattering rate as
for pointlike impurities. To gain more insight into this unusual
behavior we show in Fig.~\ref{fig:OptCondStrongImpV} the microwave
conductivity for different extensions of the scattering potential
at two of the experimentally measured frequencies, i.e., 1.14 GHz
and 13.4 GHz. Increasing the forward scattering character of the
impurity potential slightly from $V_1=V_2=V_3=100t$ to $V_1=100t$,
$V_2=90t$ and $V_3=80t$ essentially reduces the height of the peak
in the conductivity at 1.14 GHz and makes the low temperature
increase more linear. For this small deviation from isotropic
scattering, vertex corrections are small and the variation of the
conductivity can be attributed to the formation of a second
resonance in the self-energy at intermediate frequencies, see
inset in Fig.~\ref{fig:OptCondStrongImpV} (see also discussion in
Sec.~\ref{sec:SelfEnergy}). The nearly linear increase of the
self-energy below the second resonance causes the more linear
$T$-dependence of the conductivity at low temperatures. Only when
the forward scattering character of the impurity potential is
further enhanced, the vertex corrections begin to outweigh the
growing self-energy and the conductivity rises above the values
obtained for isotropic scattering. For these more extended
scattering potentials the second resonance in the self-energy
moves to lower frequencies until it merges with first resonance.
For the larger microwave frequency of 13.4 GHz, see right panel of
Fig.~\ref{fig:OptCondStrongImpV}, the anisotropy of the impurity
potential has less effect. This implies that for slightly extended
impurity potentials as considered for YBCO in
Fig.~\ref{fig:FitYBCOext} the frequency dependence becomes much
weaker and therefore a smaller concentration than in the case of
pointlike impurities is necessary to reproduce the experimentally
observed frequency dependence.
We emphasize that this analysis cannot rule out other 
explanations\cite{Hettler,DurstLeetwins} of the
quasilinear in $T$ behavior observed at low frequencies.

\section{Comparison with experimental spectra of BSCCO}
\label{sec:BSCCO}

\begin{figure}[h]
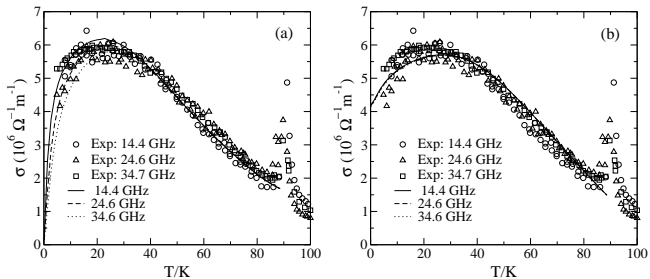

\begin{center}
\leavevmode
\begin{minipage}{.49\columnwidth}
\includegraphics[clip=true,width=.99\textwidth]{BSCCOiso.eps}
\end{minipage}
\begin{minipage}{.49\columnwidth}
\includegraphics[clip=true,width=.99\textwidth]{BSCCOExtWeak.eps}
\end{minipage}\\
\caption{Fit to the experimental microwave conductivity of BSCCO
(reproduced from Ref.~\onlinecite{leeetal}) using weak scatterers: (a)
pointlike scatterers with $V/t=1$ and $n=.049$, (b) weak extended
scatterers with $V_1/t$=2, $V_2/t$=.8, $V_3/t$=.4 and $n=.145$.
(The inelastic scattering rate has been increased by a factor of 2.6 in (a)
and 3 in (b) with respect to Ref.~\onlinecite{DSH01}).}
\label{fig:BSCCOExp1}
\end{center}
\end{figure}

In this section we want to explore what can be learned about the
type of disorder contained in the BSCCO compounds by analyzing its
microwave conductivity which has measured by S.-F. Lee {\it et
al.}~\cite{leeetal}. The temperature and frequency dependence of
the microwave conductivity in BSCCO (see symbols in
Fig.~\ref{fig:BSCCOExp1}) is quite different than in YBCO. The
absolute value of the microwave conductivity is smaller by almost
a factor of 10 indicating that BSCCO is a dirtier compound than
YBCO. This agrees well with the observation that the conductivity
of BSCCO changes noticeably only in the
THz-regime~\cite{orenstein1}, i.e., at much higher frequencies
than in YBCO. The characteristic peak in the microwave
conductivity, however, occurs at lower temperatures than in the
cleaner system YBCO contrary to predictions for strong pointlike
scatterers~\cite{HPS}. Furthermore this peak is much less
pronounced and resembles more a plateau, suggestive of the weak
scattering limit~\cite{HPS}. Finally, the conductivity does not
apparently approach the universal value $\sigma_0$ for the lowest
temperatures and frequencies measured. This might indicate the
presence of extended scatterers, which enhance the
zero-temperature and zero-frequency limit of the conductivity as
has been shown by Durst and Lee~\cite{DurstLee}.

In order to check the applicability of these scenarios for BSCCO we compare
in Fig.~\ref{fig:BSCCOExp1} respective fits to the microwave
conductivity using (i) only pointlike weak scatterers
(Fig.~\ref{fig:BSCCOExp1}(a)) and (ii) only extended weak scatterers
(Fig.~\ref{fig:BSCCOExp1}(b)). For the inelastic scattering rate 
$\tau^{-1}_{\rm inel}(T)$ we assume the same temperature-dependence as for
YBCO~\cite{DSH01} but we allow for a different prefactor in order to
account for discrepancies between YBCO and BSCCO (the exact prefactor is
stated in the figure captions).
Obviously both disorder models (i) and (ii) result in very good
fits of the experimental microwave conductivity of BSCCO. The main
difference is that the conductivity for isotropic scatterers
approaches the universal value $\sigma_0$ for $T \to 0$ whereas the conductivity
for extended scatterers (Fig.~\ref{fig:BSCCOExp1}(b)) with the
potential parameters $V_1/t=2$, $V_2/t=0.8$ and $V_3/t=0.4$ approaches
an enhanced value.
Unfortunately, it is not possible to distinguish between these two different
scenarios from the microwave conductivity alone, because there is no
experimental data available below $T=5$K.

\begin{figure}[b]
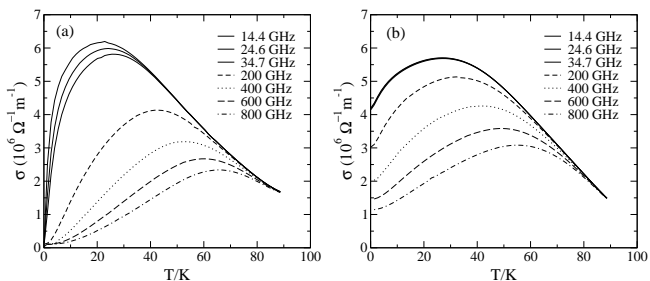

\begin{center}
\leavevmode
\begin{minipage}{.49\columnwidth}
\includegraphics[clip=true,width=.99\textwidth]{FreqIso.eps}
\end{minipage}
\begin{minipage}{.49\columnwidth}
\includegraphics[clip=true,width=.99\textwidth]{FreqExtWeak.eps}
\end{minipage}\\
\caption{Higher frequency conductivity for the disorder models of
Fig.~\ref{fig:BSCCOExp1}: (a)
pointlike scatterers with $V/t=1$ and $n=.049$, (b) weak extended
scatterers with $V_1/t$=2, $V_2/t$=.8, $V_3/t$=.4 and $n=.145$.}
\label{fig:HighFreq}
\end{center}
\end{figure}

Further insight could be gained by comparing the frequency
dependence of these two models. Impurity scattering alone would
predict a different frequency dependence for isotropic and
extended scatterers, as illustrated in Fig.~\ref{fig:HighFreq}.
Whereas the magnitude of the conductivity remains rather large at
low temperatures in the case of extended scatterers even for high
frequencies, it almost vanishes in the case of pointlike
impurities. Due to the large impurity concentration, however, this
frequency dependence is most pronounced in the THz-regime, as
observed experimentally~\cite{orenstein1}, where the contribution
of inelastic scattering cannot be neglected. Thus, the THz-data
present not only a probe of elastic impurity scattering but also
of inelastic scattering processes and are therefore not directly suitable
to distinguish between pointlike and extended impurities.

The only way we can proceed now is try to exclude one of
the two models indirectly via analyzing an additional
observable. Thus we will argue in the following that disorder
model (i) containing 4.9\% weak isotropic scatterers with a scattering
strength of $V$=1$t$ would yield an unrealistically large normal state
scattering rate.
Assuming a normal state density of states of $1/8t$ yields
a normal state scattering rate of $\tau^{-1} \approx 0.6 T_c$ for $t$=120meV.
According to Abrikosov Gorkov's
scaling law this would reduce $T_c$ by 25\%, which we
consider as an unreasonably large suppression because
$T_c \approx 93$K in the samples used for microwave conductivity in
Ref.~\onlinecite{leeetal}, which is close to the highest values of $T_c$
measured for the BSCCO-compounds.
Extended impurities, on the other hand, act mainly as small angle
scatterers and affect $T_c$ much less than isotopic
scatterers~\cite{Kee,otherTc}.
 This allows us to assess model (i),
which consists only of weak pointlike scatterers, as very unlikely
and to conclude that at least a large  fraction of the disorder in
the BSCCO compounds should be attributed to extended scatterers.
This could be confirmed by experiments on crystals at lower $T$.

So far we have focused on the effect of weak scatterers which we
attribute to the out-of plane disorder introduced by doping.
Defects whiting the CuO$_2$-planes like Zn-substitution for Cu, on
the other hand, are generally considered to act as strong
pointlike scatterers. These strong scattering defects have been
observed in STM-experiments~\cite{Hudson} on BSCCO-compounds and are
often assumed to be the main source of disorder in the
YBCO-compounds although their concentration is very low.
We therefore now address the question of whether our model for the
microwave conductivity in BSCCO is compatible with an additional
small concentration of strong pointlike impurities, which are most
likely also present in the compound used for measurements of the
microwave conductivity of BSCCO in Ref.~\onlinecite{leeetal}.

\begin{figure}[t]
\begin{center}
\leavevmode
\includegraphics[clip=true,width=.9\columnwidth]{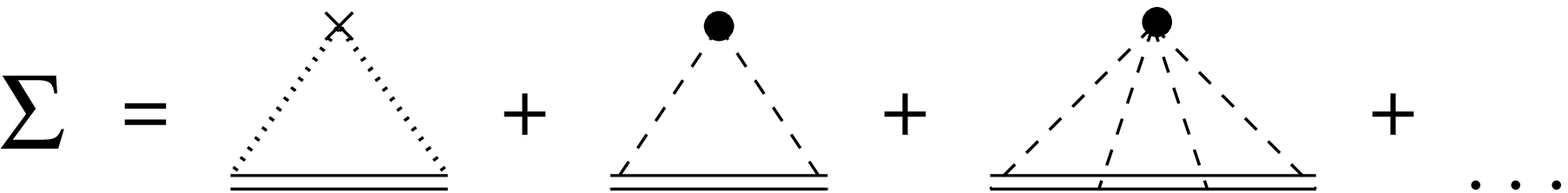}
\\[.5cm]
\includegraphics[clip=true,width=.9\columnwidth]{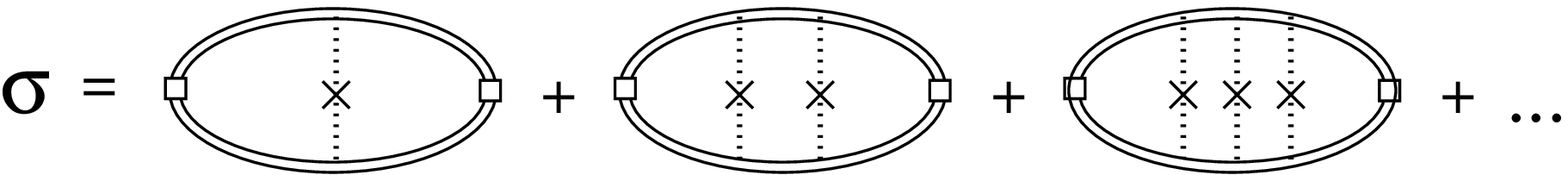}
\caption{Diagrams for the self-energy $\Sigma$ and
microwave conductivity $\sigma$ considered in the realistic disorder
model for BSCCO. Circles denote the pointlike strong impurities, which
are treated in T-matrix approximation. Crosses denote the weak
extended scatterers, which are treated in Born approximation. Only the
extended, i.e., only the weak scatterers contribute to the vertex corrections.}
\label{fig:Model2Imp}
\end{center}
\end{figure}

We incorporate this realistic disorder model, which consists of weak
extended and strong pointlike scatterers, by calculating the diagrams depicted in
Fig.~\ref{fig:Model2Imp}. 
The weak extended scatterers are treated in
Born approximation and the strong pointlike impurities in T-matrix
approximation, where as before we consider only the $\Sigma_0$-component
of the self-energy, i.e., only diagrams of the T-matrix with an even
number of impurity lines. Because vertex corrections vanish for
pointlike scatterers, only the weak extended scatterers contribute
and the vertex corrections can be calculated in Born
approximation. Thus, the microwave conductivity $\sigma(\Omega)$
is still given by Eq.~(\ref{eq:Conductivity}) with the bubble $J(\omega,\Omega)$
as in Eq.~(\ref{eq:J}) and the vertex function
$\gamma^{\rm Born}$ given in Eq.~(\ref{eq:VertexBorn}).
Only the self-energy $\Sigma_0$ has to be calculated
self-consistently as the sum of Eq.~(\ref{eq:SelfEnIso}),
Eq.~(\ref{eq:SelfEnBorn}) and the inelastic scattering rate
\begin{eqnarray}
\Sigma_0(\omega,T) &=& -\frac{n_s}{4I_G(\omega)}
\Bigl( 1-\frac{1}{1-16 I_G^2(\omega) V_s^2} \Bigr) \\
&& - n_w (V_1^2+2V_2^2+V_3^2) I_G(\omega) -i (2 \tau_{\rm inel}(T))^{-1}. \nonumber
\end{eqnarray}
Here, $n_s$ is the concentration of strong pointlike impurities with a
scattering potential $V_s$, $n_w$ is the concentration of weak
extended scatterers characterized by the potential parameters $V_1$,
$V_2$, $V_3$ and $I_G(\omega)$ is the momentum integrated single
particle Green's function given by Eq.~(\ref{eq:SingleParticleGreen}).

\begin{figure}[b]
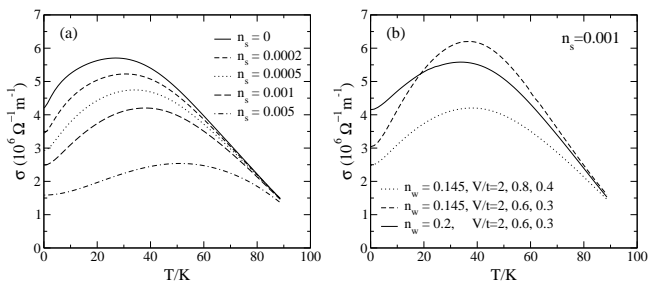

\begin{center}
\leavevmode
\begin{minipage}{.49\columnwidth}
\includegraphics[clip=true,width=.99\textwidth]{AddStrong.eps}
\end{minipage}
\begin{minipage}{.49\columnwidth}
\includegraphics[clip=true,width=.99\textwidth]{AddStrongForward.eps}
\end{minipage}\\
\caption{Effect of adding strong pointlike scatterers for the
microwave conductivity at 14.4 GHz. (a) Weak extended
scatterers with parameters of Fig.~\ref{fig:BSCCOExp1}(b), i.e.,
$n_w=0.145$ and $V_1/t=2$, $V_2/t=0.8$, $V_3/t=0.4$, and
additional strong pointlike scatterers with $V/t=100$ and different
concentrations $n_s$. (b) Fixed concentration $n_s=0.001$
of strong pointlike impurities for varying concentration
and forward scattering potential of the weak scatterers.}
\label{fig:WeakExtStrong}
\end{center}
\end{figure}

The effect of adding a small concentration of strong pointlike
scatterers to the extended weak scatterers used in
Fig.~\ref{fig:BSCCOExp1}(b) is illustrated in
Fig.~\ref{fig:WeakExtStrong}. Additional strong pointlike scatterers
mainly reduce the conductivity as can be seen
Fig.~\ref{fig:WeakExtStrong}(a). In order to raise the conductivity
to its previous values the forward scattering character of the weak
impurities must therefore be enhanced.
On the other hand, this increases the
difference between the zero-temperature value and the maximum value of the
conductivity (see dashed line in Fig.~\ref{fig:WeakExtStrong}(b))
and necessitates a larger concentration of weak
extended scatterers. Finally the lineshape of the conductivity
(solid line in Fig.~\ref{fig:WeakExtStrong}(b)) looks
similar to the one without strong pointlike impurities
but it is flatter at low temperatures than before. This poses
an upper limit for the concentration of strong scatterers compatible
with the experimental microwave conductivity of BSCCO.

\begin{figure}[t]
\begin{center}
\leavevmode
\includegraphics[clip=true,width=.85\columnwidth]{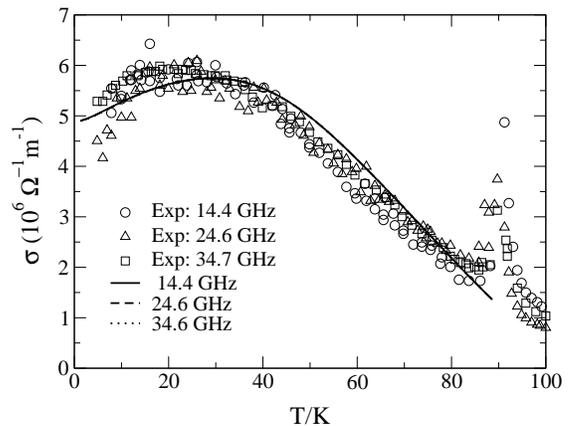}
\caption{Fit to the experimental micowave conductivity of BSCCO
(reproduced from Ref.~\onlinecite{leeetal}) using $n_w=.1$ weak extended
scatterers with $V_1/t$=3, $V_2/t$=.9, $V_3/t$=.5 and $n_s=.0005$
strong pointlike scatterers with $V_s/t=100$.
(The inelastic scattering rate has been increased by a  factor of 3.4
with respect to Ref.~\onlinecite{DSH01})}
\label{fig:BSCCOExp2}
\end{center}
\end{figure}

A fit to the experimental microwave conductivity of BSCCO containing
.05\% pointlike strong scatterers and 10\% weak extended scatterers is
shown in Fig.~\ref{fig:BSCCOExp2}. This is about the
largest concentration of pointlike strong scatterers still compatible
with the microwave conductivity. This concentration is
smaller than the .2\% observed in STM-experiments~\cite{Hudson} on BSCCO but it is very
plausible that the number of in-plane defects varies for different
compounds and maybe even between bulk and surface.

\section{Conclusions}

In summary, we have investigated the influence of extended scatterers
on the microwave conductivity of $d$-wave superconductors
by extending the approach of Durst and Lee~\cite{DurstLee}, which is
based on a nodal approximation for the quasiparticle spectrum and the
impurity potential, to finite temperatures and frequencies.
We have slightly modified the approach of Durst and Lee by considering
only vertex functions which contain an even number of impuritity lines.
This modification is necessary to ensure analyticity at finite temperatures and
frequencies when only the normal self-energy component $\Sigma_0$ is
considered.

The effect of extended scatterers on the temperature and frequency
dependence of the microwave conductivity can be summarized as
follows: for a small concentration of slightly extended strong
scatterers a second resonance forms in the self-energy at
intermediate frequencies similar to treatments which consider the
suppression of the order parameter surrounding a strong pointlike
impurity~\cite{Hettler}. This results in a more linear temperature
dependence of the conductivity at low temperatures and therefore
improves the agreement with experimental spectra of YBCO at low
temperatures.
For more extended scatterers the vertex corrections begin to
dominate over the self-energy and the magnitude of the
conductivity increases.

The microwave conductivity of BSCCO is very different compared to YBCO and
cannot be understood in terms of only strong scattering pointlike impurites. We find
that a large concentration of weak extended scatterers is necessary to
explain the observed temperature and frequency dependence of the
microwave conductivity in BSCCO, where: (i) the impurity concentration
has to be large to explain the small magnitude of the
conductivity and the negligible frequency dependence in the microwave
range, (ii) the scattering strength has to be small to account for
the plateau-like lineshape of the conductvity at small temperatures
and (iii) the range of the scattering potential has to be spatially extended
in order to keep the $T_c$-suppression reasonably small~\cite{Kee}.
Finally, we have shown that adding a small concentration of
pointlike strong scatterers, which have been observed in
STM-experiments~\cite{Hudson}, to the weak extended scatterering
potential, which we attribute to the out-of-plane disorder
introduced by doping, is still compatible with the microwave
conductivity of BSCCO. Although it would be necessary to refine
our treatment of inelastic scattering by accounting for its
frequency dependence and its contribution to vertex corrections in
order to address the conductivity in the
THz-range~\cite{orenstein1}, it is interesting to note that
elastic scattering alone would predict a very different frequency
dependence for pointlike and extended scatterers.

{\it Acknowledgments.} This work was supported by a Feodor-Lynen
Fellowship from the A. v. Humboldt Foundation (TSN) and ONR grant
N00014-04-0060 (PJH).  The authors are grateful to D.A. Bonn, R. Harris,
I. Bosovic, L.-Y. Zhu, D.~J.~Scalapino and P.~W\"olfle for
stimulating conversations.

\end{document}